\documentclass[prl,aps,twocolumn,floatfix,superscriptaddress,preprintnumbers]{revtex4-1}

\pdfoutput=1
\usepackage{mathrsfs}
\usepackage{amsfonts}
\usepackage{amsmath,amssymb,bm,bbm}
\usepackage[colorlinks=true,linkcolor=blue,citecolor=blue,urlcolor=blue]{hyperref}
\usepackage{epsfig}
\usepackage{graphicx}
\usepackage{slashed}
\usepackage{booktabs}
\usepackage{tabu}
\usepackage[dvipsnames]{xcolor}
\usepackage[normalem]{ulem}
\usepackage{tikz}
\usepackage{soul}
\usetikzlibrary{shapes,arrows}
\usetikzlibrary{positioning}
\usepackage{ulem}
\newcommand{\diff}[1]{\mathrm{d}#1}

\begin{document}
\preprint{JLAB-THY-21-3482}
\title{Global QCD Analysis of Pion Parton Distributions with Threshold Resummation}

\author{P. C. Barry}
\affiliation{Jefferson Lab,
	     Newport News, Virginia 23606, USA}
\author{Chueng-Ryong Ji}
\affiliation{Department of Physics, North Carolina State University, Raleigh, North Carolina 27695, USA \\
        \vspace*{0.2cm}
        {\bf Jefferson Lab Angular Momentum (JAM) Collaboration
        \vspace*{0.2cm} }}
\author{N. Sato}
\affiliation{Jefferson Lab,
	     Newport News, Virginia 23606, USA}
\author{W. Melnitchouk}
\affiliation{Jefferson Lab,
	     Newport News, Virginia 23606, USA}

\begin{abstract}
We perform the first global QCD analysis of pion valence, sea quark, and gluon distributions within a Bayesian Monte Carlo framework with threshold resummation on Drell-Yan cross sections at next-to-leading log accuracy.
Exploring various treatments of resummation, we find that the large-$x$ asymptotics of the valence quark distribution $\sim (1-x)^{\beta_v}$ can differ significantly, with $\beta_v$ ranging from $\approx 1$ to $> 2.5$ at the input scale.
Regardless of the specific implementation, however, the resummation induced redistribution of the momentum between valence quarks and gluons boosts the total momentum carried by gluons to $\approx 40\%$, increasing the gluon contribution to the pion mass to $\approx 40$~MeV. 
\end{abstract}

\date{\today}
\maketitle

%%%%%%%%%%%%%%%%%%%%%%%%%%%%%%%%%%%%%%%%%%%%%%%%%%%%%%%%%%%%%%%%%%%%%%%
{\it Introduction.}---\
As the lightest known hadron, the pion presents itself as a dichotomy of nature.
On the one hand, as the pseudo-Goldstone boson associated with chiral symmetry breaking it is fundamental for understanding low-energy hadronic interactions~\cite{Scherer:2012xha, Thomas:1982kv}; on the other, as a QCD bound state composed of quarks and gluons, its partonic structure reveals itself in high-energy scattering experiments much like other hadrons.
Since pions do not exist as free stationary targets from which one could scatter, such as protons or stable nuclei, details of their partonic structure have consequently been more difficult to establish empirically.
While experiments with secondary pion beams on nuclear targets~\cite{Conway:1989fs, Badier:1983mj} have provided intriguing glimpses into the pion's valence quark structure, many open questions remain.

The structure of the pion's quark distributions in the deep valence region, where a single parton carries a large fraction, $x$, of the pion's momentum, has been of considerable interest~\cite{Holt:2010vj}, particularly regarding its behavior in the limit as $x \to 1$.
An ongoing debate has pitted arguments based on perturbative QCD models, which predict an asymptotic behavior for the valence quark PDF $\sim (1-x)^{\beta_v}$ with $\beta_v=2$, against various nonperturbative models which favor smaller values $\beta_v \lesssim 1$~\cite{Ezawa:1974wm, Landshoff:1973pw, Gunion:1973ex, Farrar:1979aw, Berger:1979du, Shigetani:1993dx, Szczepaniak:1993uq, Davidson:1994uv, Hecht:2000xa, Melnitchouk:2002gh, Noguera:2015iia, Hutauruk:2016sug, Hobbs:2017xtq, deTeramond:2018ecg, Bednar:2018mtf, Lan:2019vui, Lan:2019rba, Chang:2020kjj, Cui:2020tdf, Kock:2020frx}.
There are even suggestions~\cite{Chang:2021utv} that failure to obtain $\beta_v=2$ would be in conflict with QCD itself, and the debate is motivating experimental programs aimed at discriminating between the possible large-$x$ behaviors~\cite{Arrington:2021biu}.

Recently, significant progress has been made in resumming large logarithmic corrections in perturbative QCD from partonic threshold effects in high-energy reactions through various methods and approximations \cite{Korchemsky:1992xv, Belitsky:1998tc, Moch:2005ky, Laenen:2005uz, Idilbi:2006dg, Ravindran:2006bu, Bonvini:2015ira, Bonocore:2016awd, Liu:2017pbb, Banerjee:2018vvb, Beneke:2018gvs, Hinderer:2018nkb, Bacchetta:2019tcu, Lustermans:2019cau, Dai:2021mxb, Abele:2021nyo, vanBeekveld:2021mxn}.
In particular, in a seminal study of pion-nucleus Drell-Yan (DY) lepton pair production data~\cite{Conway:1989fs, Badier:1983mj}, Aicher, Sch\"afer and Vogelsang (ASV)~\cite{Aicher:2010cb} showed that corrections from threshold resummation~\cite{Westmark:2017uig, Bonvini:2010tp} made significant contributions at large values of $x$.
A naive $\beta_v \approx 1$ behavior of the valence pion parton distribution function (PDF) in a fixed order calculation was found to yield a softer, $\beta_v \approx 2$ behavior in a resummed calculation.
Since the DY data are not very sensitive to the small-$x$ region, ASV focused on fitting the valence PDF and fixing the sea quark and gluon PDFs to those from an earlier analysis~\cite{Gluck:1999xe}.

In parallel developments, the Jefferson Lab Angular Momentum (JAM) Collaboration recently explored~\cite{Barry:2018ort} the inclusion of leading neutron (LN) electroproduction data from HERA~\cite{H1:2010hym, ZEUS:2002gig}, in addition to the DY data, to constrain the valence, sea quark and gluon distributions at low and high $x$ values, using Bayesian Monte Carlo methods.
A subsequent fixed order analysis also included high-$p_T$ DY data \cite{Cao:2021aci}, together with $p_T$-integrated data.
Other phenomenological analyses have utilized DY and prompt photon data to constrain pion PDFs \cite{Owens:1984zj, Aurenche:1989sx, Sutton:1991ay, Gluck:1991ey, Gluck:1999xe, Wijesooriya:2005ir, Novikov:2020snp}, and the growing number of recent lattice calculations~\cite{Sufian:2019bol, Izubuchi:2019lyk, Joo:2019bzr, Sufian:2020vzb, Karthik:2021qwz, Alexandrou:2021mmi, Fan:2021bcr}, some including threshold resummation~\cite{Gao:2021hxl}, is a testament to the importance of better understanding the pion's PDFs.

In this Letter we bring these strands together to perform a global QCD analysis of DY and LN data within the JAM framework that includes for the first time a systematic study of soft gluon resummation effects on pion PDFs at next-to-leading logarithmic (NLL) accuracy.
In particular, we critically examine the universality of the resummation impact on the effective $\beta_v$ parameter, by considering several 
viable resummation prescriptions, 
including the traditional Mellin-Fourier (MF) method~\cite{Mukherjee:2006uu, Bolzoni:2006ky, Aicher:2010cb, Bonvini:2010tp} and the more recently developed double Mellin (DM) method~\cite{Westmark:2017uig}.

%%%%%%%%%%%%%%%%%%%%%%%%%%%%%%%%%%%%%%%%%%%%%%%%%%%%%%%%%%%%%%%%%%%%%%%
{\it Threshold resummation of Drell-Yan.}---\
In the inclusive pion-induced DY process \cite{Drell:1970wh}, a pion beam  incident on a nuclear target, with total center of mass energy $\sqrt{S}$, produces an inclusive $\mu^+\mu^-$ pair with invariant mass $Q$.
The cross section differential with respect to $\tau=Q^2/S$ and rapidity, $Y$, can be factorized into a convolution of perturbatively calculable hard scattering coefficients $C_{ij}$ and the collinear pion and nuclear PDFs~\cite{Collins:1983ju},
\begin{multline}
    \frac{\diff^2\sigma}{\diff\tau \diff Y} 
    = \frac{4\pi\alpha^2}{9\tau S^2} \sum_{ij} \int\!\diff z\!\int\!\diff y ~C_{ij}(z,y,\mu/Q) 
    \\
    \times f_i^\pi(x_\pi,\mu)\, f_j^A(x_A,\mu),
    \label{eq.DYxsec}
\end{multline}
where the PDFs $f_{i(j)}^{\pi(A)}$ for parton flavor $i(j)$ in the pion (nucleus) are functions of the parton's light-front momentum fraction $x_{\pi(A)}$ with respect to the parent hadron, and $\mu$ is the factorization scale.
The integration variables are 
    $y=(u-z)/[(1-z)(1+u)]$, 
where
    $u=e^{-2Y} (x_\pi/x_A)$,
and
    $z=Q^2/\hat{s}$,
with $\hat{s}=x_\pi x_A S$ the partonic invariant mass squared.

At leading order (LO), only the $q\bar{q}$ channel contributes, and the 
hard coefficient,
    $C_{q\bar{q}}^{(0)} = \delta(1-z)$, 
allows the PDFs to be evaluated at momentum fractions $x_{\pi,A}^0=\sqrt{\tau} e^{\pm Y}$.
At next-to-leading order (NLO), the tree-level $qg$ channel enters, while the $gg$ and $qq'$ channels appear at next-to-next-to-leading order (NNLO).
Focusing on the $q\bar{q}$ channel, the hard coefficients can be schematically organized as a sum of a Born term that describes the LO $q\bar{q}$ annihilation, a virtual term that includes interference of the one-loop and Born diagrams, and a real emission term describing the radiation in the final state.
The $q\bar{q}$ channel plays the dominant role, and is the focus of this work.

The phase space for real gluon radiation vanishes in the limit $z \to 1$.
In the infrared region where the real gluon radiation becomes soft, mass singularities from propagators in the real and virtual diagrams cancel up to terms of the form $\alpha_s^k \big( \log^{2k-1}(1-z) \big)/(1-z)$, known as threshold logarithms, which appear order by order in perturbation theory.
At NLO, for example, in the rapidity-integrated DY cross section the hard coefficient includes terms 
$C_{q\bar{q}}^{(1)}(z) \propto \alpha_s \big(\log(1-z)\big)/(1-z)$.

Because the PDFs are steeply falling at large $x$, the partonic cross section in the threshold region
plays a substantial role in the overall hadronic cross section~\cite{Aicher:2010cb, Westmark:2017uig, Shimizu:2005fp, Catani:1996yz}.
Changes in the perturbative $C_{ij}$ coefficients are compensated by changes in the PDFs to produce the same physical cross section~\cite{Aicher:2010cb}.
Near the partonic threshold, the logarithms become increasingly important, and must be resummed in order to maintain the integrity of the perturbative expansion.
The threshold resummation framework of Sterman~\cite{Sterman:1986aj} and Catani and Trentadue \cite{Catani:1989ne} allows one to systematically include the large logarithmic contributions to the DY process from soft gluon emissions to all orders of~$\alpha_s$.

The calculation of threshold resummation is not straightforward in momentum space.
In Mellin space, however,
the phase space integrals of multiple soft gluon emissions decouple, allowing convenient organization of the threshold logarithms to be resummed.
For the rapidity-dependent DY cross section, an additional transformation is needed beyond the single Mellin, which we take as either an additional Fourier transform in~$Y$~\cite{Sterman:2000pt, Mukherjee:2006uu} or a double Mellin transform in $x_\pi^0$ and $x_A^0$~\cite{Westmark:2017uig},
\begin{subequations}
\label{eq.momentDYxsec}
\begin{align}
&\sigma_{\rm \mbox{\tiny MF}}(N,M) 
\equiv 
\int_0^1 \diff\tau \tau^{N-1}\!
\int_{\log \sqrt{\tau}}^{\log \frac{1}{\sqrt{\tau}}} \diff Y
e^{i M Y} \frac{\diff^2\sigma}{\diff\tau \diff Y}, 
\\
&\sigma_{\rm \mbox{\tiny DM}}(N,M) \equiv 
\int_0^1 \diff x_\pi^0\, (x_\pi^0)^{N-1} 
\int_0^1 \diff x_A^0\, (x_A^0)^{M-1} \frac{\diff^2\sigma}{\diff\tau \diff Y}.
\end{align}
\end{subequations}
The convolution form of the cross section~(\ref{eq.DYxsec}) decouples in conjugate space into
a simple product of hard coefficients and PDFs.
In the {\small MF} case, defining a partonic rapidity
    $\widehat{Y} = Y - \frac12 \log(x_\pi/x_A) = -\frac12\log u$
allows the partonic cross section to be written as a Mellin transform in $z$ and Fourier transform in $\widehat{Y}$.
Note that the
threshold region $z \approx 1$ maps into large $N$ in Mellin space.

The formal definition~\cite{Catani:1989ne, Catani:1996yz} of the Mellin transform for the resummed hard coefficients implies an evaluation at the Landau pole.
Although the Landau pole can be found explicitly in the resummed coefficients as a function of $N$, there is ambiguity in how it should be treated~\cite{Catani:1996yz, Bonvini:2010tp, Westmark:2017uig, Bauer:2000yr}.
The large logarithms being resummed appear at 
    $z \leq 1-1/(N e^{\gamma_E})$,
where $\gamma_E$ is the Euler constant, truncating the full Mellin transform and giving rise to analytic terms~\cite{Aicher:2010cb, Westmark:2017uig}.
Since the large Mellin logarithms are far from the Landau pole, $N_L$, the minimal prescription {\small (MP)}~\cite{Catani:1996yz} makes use of special contours $C^{\rm \mbox{\tiny MP}}$ that avoid $N_L$, while enclosing poles from the PDFs in the Mellin inversion.
To use the Mellin expressions for the threshold resummation we solve DGLAP evolution in Mellin space, which 
avoids numerical instability of the hard coefficients in an $x$-space computation~\cite{Westmark:2017uig, Bonvini:2010tp}.

To compare with experiment, we invert Eqs.~(\ref{eq.momentDYxsec}) to momentum space to obtain the differential cross section,
\begin{equation}
\begin{split}
\frac{\diff^2\sigma}{\diff\tau \diff Y}
&=\int_{-\infty}^{\infty} \frac{\diff M}{2\pi} e^{-i M Y} 
  \int_{C^{\rm MP}_N} \frac{\diff N}{2\pi i}\,
  \tau^{-N} \sigma_{\rm \mbox{\tiny MF}}(N,M)
\\
&=\int_{C_M^{\rm MP}} \frac{\diff M}{2\pi i} (x_A^0)^{-M}
  \int_{C_N^{\rm MP}} \frac{\diff N}{2\pi i} (x_\pi^0)^{-N} 
  \sigma_{\rm \mbox{\tiny DM}}(N,M).
\label{eq.DYxsecinversions}
\end{split}
\end{equation}
Since the resummed expansion is performed at NLL, the threshold logarithms at NLO also appear in the resummed calculation.
Expanding the NLL terms in orders of $\alpha_s$ and subtracting up to the $\mathcal{O}(\alpha_s)$ terms ensures that the NLO pieces are not double counted.

In addition to the difference between the {\small MF} and {\small DM} methods, another ambiguity appears in the {\small MF} case.
The real emission terms proportional to the threshold logarithms include a factor $\delta\big(\widehat{Y} \pm \log(1/\sqrt{z})\big)$, which after the Fourier transform produces a $\cos(M\log(1/\sqrt{z}))$ term.
Expanding the cosine reveals $\mathcal{O}((1-z)^2)$ corrections, which are subleading near threshold, and the cosine may be approximated as 1~\cite{Bolzoni:2006ky}.
In this work, we refer to this as the ``expansion'' method.

Alternatively, the cosine may be kept without expanding~\cite{Mukherjee:2006uu, Aicher:2010cb}, which we refer to as the ``cosine'' method.
Similar subleading $\mathcal{O}((1-z)^n)$ terms appear in non-$q\bar{q}$ channels, so one may argue that a consistent treatment requires inclusion of the other channels.
On the other hand, because of its dependence on $M$, the cosine method provides additional information about the rapidity distribution.
In our analysis, we systematically explore all three methods, and quantify the dependence of the fitted PDFs on the resummation method choices.

%%%%%%%%%%%%%%%%%%%%%%%%%%%%%%%%%%%%%%%%%%%%%%%%%%%%%%%%%%%%%%%%%%%%%%%
{\it Bayesian inference.}---\
Our numerical analysis uses the Bayesian Monte Carlo methodology developed by the JAM Collaboration~\cite{Sato:2016tuz, Sato:2016wqj, Ethier:2017zbq, Lin:2017stx, Sato:2019yez, Moffat:2021dji, Cammarota:2020qcw, Bringewatt:2020ixn}, whereby the probability of the set $\bm{a}$ of best fit parameters conditioned on the data is proportional to the likelihood of the data given the parameter set.
We sample the posterior probability distribution
    $\mathcal{P}(\bm{a}|{\rm data})
    \propto \mathcal{L}({\rm data}|\bm{a}) \pi(\bm{a})$, 
where 
    $\mathcal{L}({\rm data}|\bm{a})
    = \exp{(-\frac12\chi^2(\bm{a},{\rm data}))}$
is the likelihood function and $\pi(\bm{a})$ is the prior distribution.
Data resampling is used 
with Gaussian noise added to the data within the quoted uncorrelated uncertainties~\cite{Sato:2016tuz, Sato:2016wqj}, and the posterior distribution is sampled via multiple likelihood regressions.
The expectation value E and variance V of an ``observable'' $\mathcal{O}$ (such as a PDF or cross section) are computed from
E$[\mathcal{O}]=(1/N_{\rm rep})\sum_i \mathcal{O}(\bm{a}_i)$ and
V$[\mathcal{O}]=(1/N_{\rm rep})\sum_i \big({\rm E}[\mathcal{O}]-\mathcal{O}(\bm{a}_i)\big)^2$,
where the sum is over $N_{\rm rep}$ number of replicas.

The pion PDF for a given flavor $i$ is parametrized at the input scale, $\mu_0 = m_c = 1.27$~GeV, by the form
\begin{equation}
f_i(x,\mu_0;\bm{a}_i)
= N_i\, x^{\alpha_i}(1-x)^{\beta_i}(1+\gamma_i x^2),
\label{eq.PDFparams}
\end{equation}
for the set of parameters $\bm{a}_i=\{N_i,\alpha_i,\beta_i,\gamma_i\}$.
We assume charge symmetry for the valence quark PDF,
    $q_v \equiv \bar{u}_v^{\pi^-}\!
    = \bar{u}^{\pi^-}\!-u^{\pi^-}\!
    = d_v^{\pi^-}$,
and a flavor symmetric sea,
    $q_s \equiv u^{\pi^-}\!
    = \bar{d}^{\pi^-}\!
    = s^\pi = \bar{s}^\pi$.
Valence quark number conservation,
    $\int_0^1 \diff x\, q_v(x,\mu) = 1$,
and the momentum sum rule,
    $\int_0^1 \diff x\, x\,
    \big( 2 q_v(x,\mu) + 6 q_s(x,\mu) + g(x,\mu) \big) = 1$,
constrain the normalizations $N_v$ and $N_s$, respectively.
We further set $\gamma_{s,g} = 0$, as these could not be constrained by existing data, and choose
$N_g>0$ and $\gamma_v > -1$ to avoid PDFs becoming negative.

The pion valence, sea quark, and gluon PDFs are fitted to the available pion-nucleus Drell-Yan data from E615~\cite{Conway:1989fs} (61 data points) and NA10 \cite{Badier:1983mj} (56 points), presented as
    $\diff\sigma/\diff x_F\, \diff\sqrt{\tau}$, 
where 
    $x_F \equiv x_\pi^0-x_A^0$.
The analysis is limited to the range
    $4.16 < Q < 7.68$~GeV and $0 < x_F < 0.9$
to avoid the $J/\psi$ and $\Upsilon$ resonances and edges of phase space.
For the nuclear PDFs we use the Eskola \textit{et al.} parametrization~\cite{Eskola:2016oht}, although using the
nCTEQ fit~\cite{Kovarik:2015cma} showed no differences in final results.
As in previous analyses~\cite{Barry:2018ort, Cao:2021aci}, we include leading neutron electroproduction data at small $x$ from HERA (58 points for H1~\cite{H1:2010hym} and 50 points for ZEUS~\cite{ZEUS:2002gig} data), which at very forward angles are expected to be dominated by pion exchange~\cite{Sullivan:1971kd, McKenney:2015xis}.

%%%%%%%%%%%%%%%%%%%%%%%%%%%%%%%%%%%%%%%%%%%%%%%%%%%%%%%%%%%%%%%%%%%%%%%
{\it Monte Carlo analysis.}---\
Fairly good fits to the data can be obtained for all resummation prescriptions considered, including no resummation. 
For the fixed order NLO analysis, a total $\chi^2$ per datum was found of $\chi^2_{\rm dat}=0.81$ for the combined DY and LN datasets, consistent with the recent analysis~\cite{Cao:2021aci}, while including resummation with the cosine, expansion, and double Mellin methods gives a total $\chi^2_{\rm dat}$ of 1.29, 0.95, and 0.80, respectively.
The $\chi^2_{\rm dat}$ variation comes almost exclusively from the DY datasets, with the low-$x$ LN data mostly unaffected by the perturbative QCD treatment of the DY cross sections.

\begin{figure}[t]
    \centering
    \includegraphics[width=0.5\textwidth]{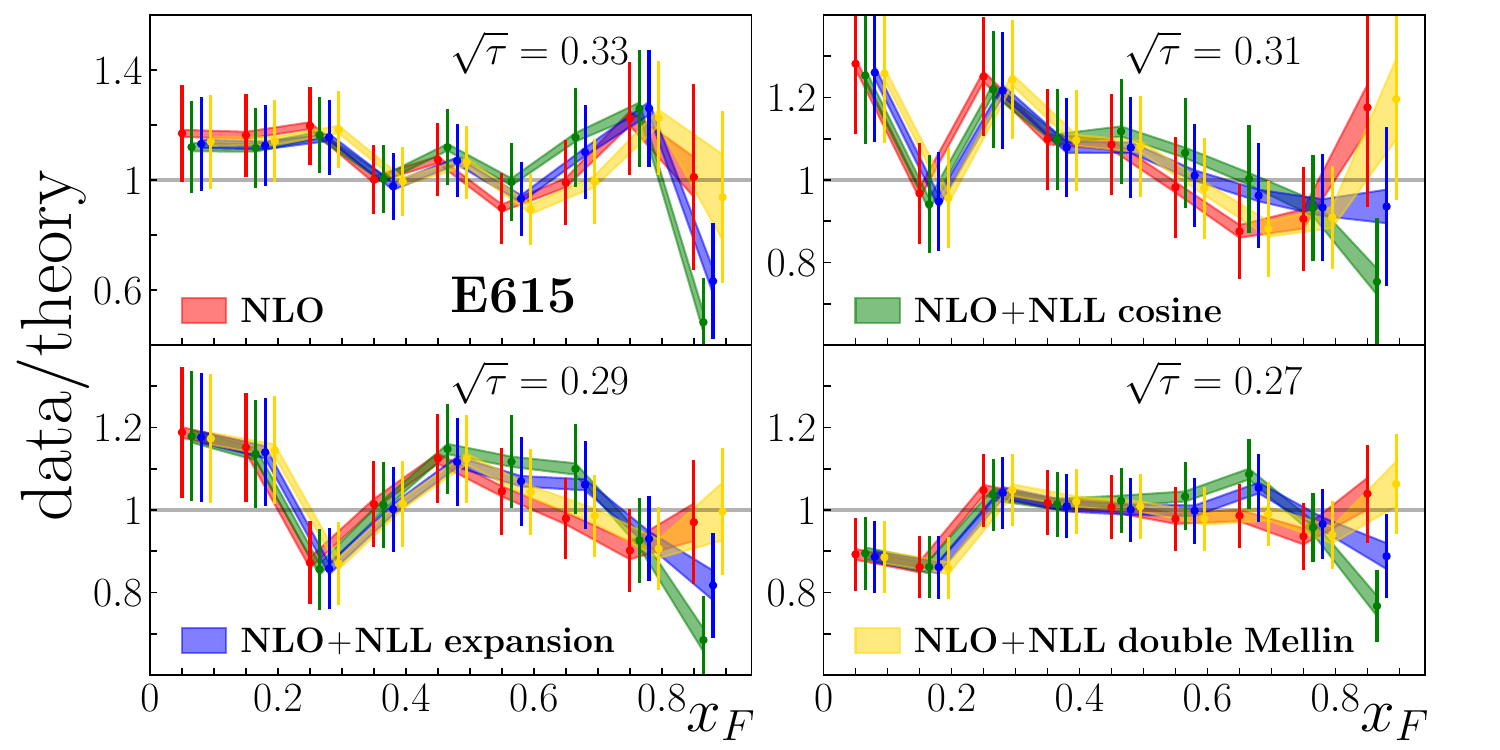}
    \caption{Ratios of several representative bins of E615 DY data~\cite{Conway:1989fs} to the calculated cross sections versus $x_F$ at fixed $\tau$ for the NLO fixed order (red) and NLO+NLL cosine (green), expansion (blue), and double Mellin (gold) formulations, with 1$\sigma$ uncertainty bands.}
    \label{f.dvt}
\end{figure}

\begin{figure*}[t]
    \includegraphics[width=0.9\textwidth]{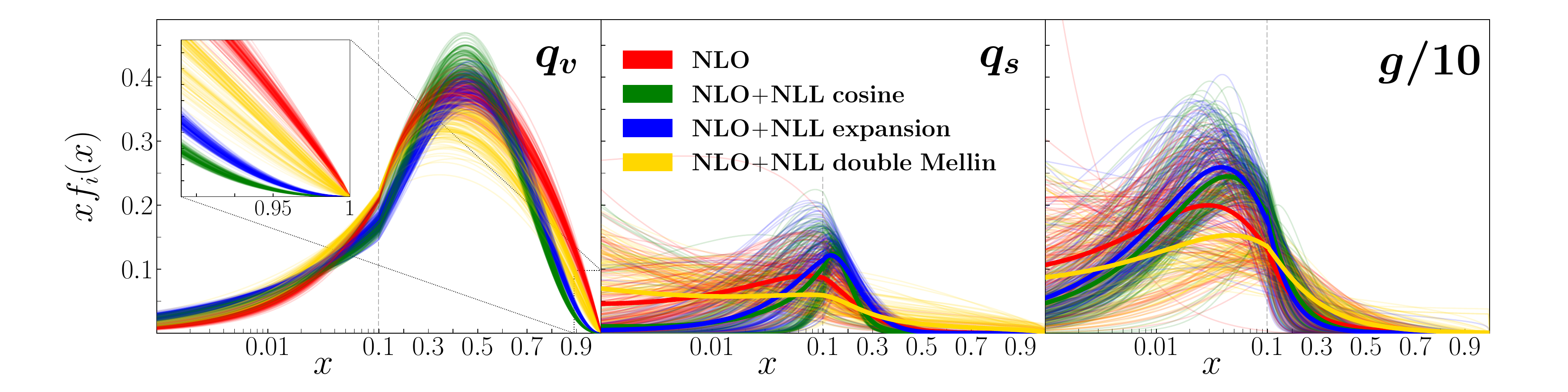} \vspace*{-0.4cm}
    \caption{Distribution of replicas for the pion valence quark {\bf (left)}, sea quark {\bf (middle)}, and gluon ($\times 1/10$) {\bf (right)} PDFs versus $x$ at the scale $\mu_0$ for the NLO fixed order (red), and NLO+NLL cosine (green), expansion (blue), and double Mellin (gold) analyses.
    The inset in the left panel magnifies the very large-$x$ region.
    The central values of the sea quark and gluon posterior samples are indicated by solid lines.}
    \label{f.PDFs}
\end{figure*}

The ratio of the experimental DY data to the calculated cross sections is shown in Fig.~\ref{f.dvt} versus $x_F$ in bins of $\sqrt{\tau}$, for a representative subset of the E615 data~\cite{Conway:1989fs}.
The comparison of the fixed order NLO fit with those with resummation shows most variance of the data descriptions in the large-$x_F$ region, while at small $x_F$ no discernable differences are seen among the prescriptions.
The NLO and double Mellin methods give similar descriptions of the data, while the cosine and expansion methods have larger contributions to the theory at high $x_F$, rendering a ratio below unity.
Based on the $\chi^2_{\rm dat}$ criteria, we conclude that the NLO, NLL expansion and double Mellin prescriptions give equally good fits, while the cosine method is slightly disfavored.

The fitted valence quark, sea quark, and gluon (scaled by a factor 1/10) PDFs are shown in Fig.~\ref{f.PDFs} for the various resummation prescriptions at the input scale $\mu=\mu_0$.
The extracted valence distributions with NLL resummation
all clearly show a softer falloff as $x \to 1$, as magnified in the inset.
The NLO valence PDF has the hardest distribution, followed by the double Mellin method, while the expansion method is softer, and the cosine method yields the softest falloff.
As seen in Fig.~\ref{f.dvt}, the cosine method tends to overpredict the data at large $x_F$, indicating that the contribution to the hard coefficients becomes too large and the PDFs cannot sufficiently adjust because of restrictions from the sum rules and LN data.
To compensate for the large hard coefficients, the PDF becomes suppressed to bring the 
calculated cross section down to match the data.
Interestingly, when positivity of the PDFs is not enforced, the cosine method admits slightly negative valence solutions at $x \gtrsim 0.85$.

For the sea quark and gluon PDFs, there is obviously greater spread, with a difference that the double Mellin method admits a sea quark shape that
is somewhat larger at high $x$ than the other methods.
Additionally, the central values exhibit larger gluon distributions for the cosine and expansion methods compared to the NLO for $0.01 \lesssim x \lesssim 0.1$, whereas the double Mellin resummation favors a larger gluon at higher $x$.

\begin{figure}[b]
    \includegraphics[width=0.40\textwidth]{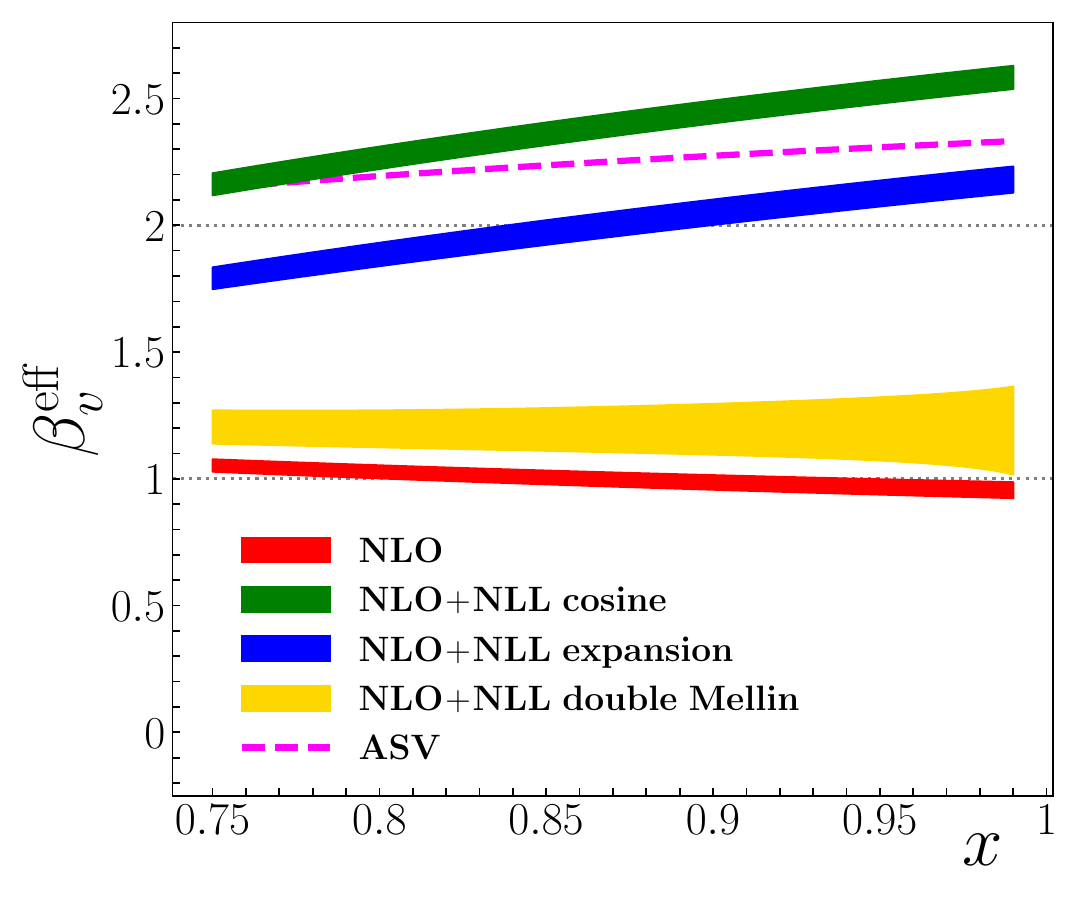} \vspace*{-0.2cm}
    \caption{Effective exponents $\beta_v^{\rm eff}$ for the various prescriptions versus $x$ at the scale $\mu_0$, compared with the ASV extraction~\cite{Aicher:2010cb}.
    The values $\beta_v^{\rm eff}=1$ and 2 are shown for reference.}
    \label{f.beff}
\end{figure}

To quantify the behavior of the valence PDF at large~$x$, we compute the effective $\beta_v$ parameter reflecting the exponent of the $(1-x)^{\beta_v}$ term in Eq.~(\ref{eq.PDFparams}), defined by~\cite{Nocera,Courtoy:2020fex,Courtoy:2021xpb}
\begin{equation}
\beta_v^{\rm eff}(x,\mu) = \frac{\partial\log\big|q_v(x,\mu)\big|}{\partial \log(1-x)}.
    \label{eq.beff}
\end{equation}
The results for the various resummation scenarios are shown in Fig.~\ref{f.beff} as a function of $x$ at the input scale, $\mu_0$,
and compared with the ASV~\cite{Aicher:2010cb} analysis that fit the valence PDF to DY data using the cosine method of threshold resummation.
In contrast to this work, ASV set $\langle x \rangle_v$ between 0.55 and 0.7 and fixed the sea quark and gluon distributions from the earlier GRS analysis~\cite{Gluck:1999xe}.
Consistent with previous studies~\cite{Sutton:1991ay, Barry:2018ort, Novikov:2020snp, Cao:2021aci}, our NLO analysis shows a linear falloff of the valence PDF with $\beta_v^{\rm eff} \approx 1$ for $x \to 1$.
Inclusion of threshold resummation results in a wide variety of $\beta_v^{\rm eff}$ values, with the cosine and expansion methods yielding $\beta_v^{\rm eff} > 2$, consistent with ASV~\cite{Aicher:2010cb}, and as large as $\approx 2.6$.

On the other hand, with the DM method the effective exponent is much closer to the NLO case, with $\beta_v^{\rm eff} \approx 1.2$ as $x \to 1$.
In fact, a recent study~\cite{Lustermans:2019cau} pointed out problems with the MF methods neglecting some leading power effects, which are, in contrast, able to be accommodated with the DM method.
This would suggest that $\beta_v^{\rm eff}$ values $\sim 1$ are preferred in the more consistent DM approach, although it would be desirable to have empirical confirmation of this with additional observables sensitive to pion PDFs at large $x$.

\begin{table}[b]
\centering
\caption{Total momentum fractions of the valence quark, sea quark, and gluon distributions at the input scale $\mu=\mu_0$ for various resummation prescriptions.}
\begin{tabular}{lccc}
\hline
~Resummation method      & ~~$\langle x \rangle_v$~~ & ~~$\langle x \rangle_s$~~ & ~~$\langle x \rangle_g$~~ \\
\hline
~NLO                     & ~0.53(2)~ & ~0.14(4)~ & ~0.34(6)~ \\
~NLO+NLL cosine          & ~0.47(2)~ & ~0.14(5)~ & ~0.39(6)~ \\
~NLO+NLL expansion       & ~0.46(2)~ & ~0.16(5)~ & ~0.38(6)~ \\
~NLO+NLL double Mellin~~~& ~0.46(3)~ & ~0.15(7)~ & ~0.40(5)~ \\
\hline
\end{tabular}
\label{t.moms}
\end{table}

{\it Momentum fractions and pion mass decomposition.}---\
A consequence of applying the NLL corrections to the DY cross section is that the large-$x$ momentum of the valence quarks is redistributed to gluons at small $x$.
The values of the total momentum fractions 
    $\langle x \rangle_i \equiv \int_0^1 \diff x x f_i(x)$ 
for the different flavors are shown in Table~\ref{t.moms}.
Interestingly, while the shapes of the PDFs for the various resummed fits differ, the momentum fractions are rather stable, with $\approx 5$\%--6\% of the momentum moving from the valence quark to the gluon sectors.

This has important implications for the decomposition of the pion mass into the quark and gluon energy and momentum, and trace anomaly contributions~\cite{MassDecomposition}.
In particular, the gluon contribution to the mass is given by 3/4 of its momentum fraction, which amounts to 40(6)~MeV, or $\approx 30\%$ of the pion mass.
This represents an increase of $\approx 14\%$ on the gluonic fraction of the mass from the NLO analysis without resummation.

%%%%%%%%%%%%%%%%%%%%%%%%%%%%%%%%%%%%%%%%%%%%%%%%%%%%%%%%%%%%%%%%%%%%%%%
{\it Outlook.}---\
In the future, theoretical improvements will extend the treatment of resummation to NNLO corrections, allowing the analysis to be generalized to the $qg$, $gg$ and $qq'$ channels and resummation effects on sea quark and gluon PDFs \cite{Beneke:2018gvs,Lustermans:2019cau}.
Concurrently, planned high luminosity tagged deep-inelastic scattering experiments at Jefferson Lab~\cite{TDISProposal} and the future Electron-Ion Collider~\cite{AbdulKhalek:2021gbh} on leading proton and neutron production at kinematics complementary to HERA will help isolate pion exchange contributions to valence and sea quarks, and test the Sullivan mechanism~\cite{Sullivan:1971kd} in the low-mass region of the pion structure function.

The proposed COMPASS++/AMBER~\cite{COMPASS} experiment at CERN to measure pion-nucleus DY cross sections, including its $p_T$ dependence, would allow the large-$x$ region to be further probed, providing a vital check on the Fermilab DY data and sensitivity to different nuclear targets.
The availability of kaon beams would also allow global QCD analysis to elucidate for the first time the kaon's quark and gluon structure.\\

%%%%%%%%%%%%%%%%%%%%%%%%%%%%%%%%%%%%%%%%%%%%%%%%%%%%%%%%%%%%%%%%%%%%%%%
{\it Acknowledgments.}---\
We thank F.~Ringer for useful discussions.
This work was supported by the US Department of Energy (DOE) contract DE-AC05-06OR23177, under which Jefferson Science Associates, LLC operates Jefferson Lab, by the US DOE contract DE-FG02-03ER41260, and by the US DOE, Office of Science, Office of Workforce Development for Teachers and Scientists, Office of Science Graduate Student Research (SCGSR) program. 
The SCGSR program is administered by the Oak Ridge Institute for Science and Education for the DOE under contract number DE‐SC0014664. 
The work of NS was supported by the DOE, Office of Science, Office of Nuclear Physics in the Early Career Program.

%%%%%%%%%%%%%%%%%%%%%%%%%%%%%%%%%%%%%%%%%%%%%%%%%%%%%%%%%%%%%%%%%%%%%%%

\end{document}